\documentstyle[12pt]{article}
\renewcommand\theequation{\thesection.\arabic{equation}}
\begin{document}
\setlength{\unitlength}{1mm}
\newcommand{\te}{\theta}
\newcommand{\bee}{\begin{equation}}
\newcommand{\ene}{\end{equation}}
\newcommand{\tra}{\triangle\theta_}

{\hfill Preprint JINR E2-93-291} \vspace*{2cm} \\
\begin{center}
{\Large\bf The Heat Kernel Expansion on a Cone \\}
\end{center}

\begin{center}
{\Large\bf and Quantum Fields Near Cosmic Strings}
\end{center}

\bigskip\bigskip

\begin{center}
{\bf D.V. Fursaev}
\footnote{e-mail: fursaev@theor.jinrc.dubna.su}
\end{center}

\begin{center}
{\it Laboratory of Theoretical Physics, Joint Institute for\\
Nuclear Research, Head Post Office, P.O.Box 79, Moscow,\\
Russia}
\end{center}

\bigskip\bigskip\bigskip

\begin{center}
{\it PACS number(s): 03.70+k, 11.17.+y, 04.60.+n }
\end{center}

\bigskip

\begin{center}
{\it Short title}: Quantum Fields near Cosmic Strings
\end{center}

\begin{abstract}
An asymptotic expansion of the trace of the heat kernel on a cone where
the heat coefficients have a delta function behavior at the apex is obtained.
It is used to derive the renormalized effective action and total
energy of a self-interacting quantum scalar field on the cosmic string
space-time. Analogy is pointed out with quantum theory
with boundaries. The surface infinities in the effective action are shown
to appear and are removed by renormalization of the string tension. Besides,
the total renormalized energy turns out to be finite due to cancelation
of the known non-integrable divergence in the energy density of the field
with a counterterm in the bare string tension.

\end{abstract}

\vspace*{2cm}

\newpage
\baselineskip=.8cm

\section{Introduction}
\setcounter{equation}0

In the last years a considerable number of works was devoted to the quantum
theory on space times with conical singularities \cite{a5}-\cite{a10}.
The line element on a conical space can be written in a form like on the
plane in polar coordinates
\begin{equation}
ds^2=dr^2 + r^2d\varphi ^2~~,~~~~0\leq r\leq \infty
\label{1.1}
\end{equation}
but with a polar angle $\varphi$ ranging from $0$ to an arbitrary
positive parameter $\alpha$. Besides, the cone (\ref{1.1}) can be considered
as a space whose curvature is completely concentrated at the apex $r=0$ and
looks like a delta function \cite{a3}.
Singularities of that sort arise at the points on the world sheet of idealized
cosmic strings with zero thickness \cite{a3},\cite{a4}. Even if the string
space-time is flat out the world sheet, its topology is non-trivial and
therefore the
spectrum of vacuum fluctuations gets modified as compared to the case of the
Minkowsky space.  This effect has been investigated by many authors
\cite{a13}-\cite{a19} who have determined the expectation value of the
renormalized energy momentum tensor.  Also of interest is quantum theory on
orbifold factors of the Riemannian manifolds \cite{a7}-\cite{a10} where
conical singularities appear at fixed points of the corresponding isometry
groups.

In the present paper we investigate the global effects of vacuum  polarization
around a cosmic string by using the trace of the heat kernel on the
cone (\ref{1.1}) that is shown to look essentially different at
asymptotically small values of the proper time as compared to the plane heat
kernel.
For this reason, the effective action obtained on its base includes a
surface divergent functional given on the string world sheet. It is
interesting that these surface infinities can be removed by
renormalization of the string tension rendering finite the total renormalized
energy.

This indicates a close analogy with quantum theory on manifolds with
boundaries \cite{a20},\cite{a23} where similar divergent terms appear on
boundary surfaces giving rise to renormalization of bare surface
gravitational actions.
The analogy can be continued further to
demonstrate that the total renormalized energy is finite owing to
cancellation of the non-integrable divergence in the energy density with
a surface counterterm resulting from the bare string tension.

The remainder of this paper is organized as follows. In Section 2, an
asymptotic expansion of the trace of the heat kernel on a cone in powers of
the proper time is found. To characterize the effect of singularity at the
apex, a more general problem is worth to be set up. In its framework the
diagonal part of the kernel is considered as a functional and the heat
coefficients turn out to have a delta function behavior at the cone tip. It is
used in Section 3 to derive the renormalized effective action, including a
surface term, and the total energy of a self-interacting scalar field
around infinitely thin straight string. The approach by Critchley, Dowker and
Kennedy \cite{a20} is explored then in Section 4 to reconcile our result with
calculations \cite{a13}-\cite{a19} that have demonstrated a non-integrable
character of the renormalized energy density.  Conclusions and remarks are
presented in Section 5.

The effects of the curvature are partially considered in Appendix A for the
case of the sphere with two conical singularities at its poles and some exact
results concerning the generalized zeta-function are presented.

\section{The heat kernel on a cone}
\setcounter{equation}0

The heat kernel $K_{\alpha}$ of the Laplace-Beltrami operator
$\triangle_{\alpha}$ on the cone (\ref{1.1}) is a solution of the
Schr\"odinger-like equation
\begin{equation}
\left(\partial/\partial s +\triangle_{\alpha}(x)\right)K_{\alpha}(x,x',s)=0
\label{2.1}
\end{equation}
with the boundary condition
$$
K_{\alpha}(x,x',0)=\delta_{\alpha}(x,x')~~~~.
$$
($\delta_{\alpha}$ is the delta function on (\ref{1.1})).

Let us define the diagonal part of the heat kernel $K_{\alpha}(x,x,s)$ as
a functional on the functions $f(r,\varphi)$ integrable on the cone and such
that the product
$rf(r,\varphi)$ is an infinitely differentiable function at zero radius
$r=0$.
Then, as it will be shown below, the following expansion
\begin{equation}
Tr\left(e^{-s\triangle_{\alpha}}f\right)\equiv\int\sqrt{g(x)}~d^2x~
K_{\alpha}(x,x,s)f(x)= {1 \over 4\pi s}
\sum_{n=0}^{\infty}a_{\alpha,n}(f)s^{n/2}+ES~~~
\label{3.1}
\end{equation}
as $s\rightarrow 0$ holds, where
$\sqrt{g(x)}~d^2x=rdrd\phi$ is the integration measure on the cone (\ref{1.1}),
$a_{\alpha,n}(f)$ are functionals on the chosen space of functions and $ES$
means the
terms that vanish exponentially as $s\rightarrow 0$.
It is interesting
to mention that this series includes half-integer powers of the
proper time $s$ and has therefore the form of an expansion on a
two-dimensional manifold with a boundary \cite{b1}.

The kernel
$K_{\alpha}$ can be constructed explicitly if the eigenfunctions and
eigenvalues of the Laplace-Beltrami operator on the cone are known. We shall
consider the Friedrichs self-adjoint extension of the operator
$\triangle_{\alpha}$ so far as in this case it is positive. This corresponds
to the wave functions regular at the conical singularity \cite{a11}.

The solution of the problem (\ref{2.1}) can be given then in an integral form
by using the heat kernel $K_{2\pi}\equiv K$ of the Laplace operator
$\triangle$ on the
plane \cite{a5},\cite{a12}
\begin{equation}
K_{\alpha}\left(x,x',s\right)={i
\over 2\alpha} \int_{C} \cot \left(\pi \alpha^{-1} w\right)
K\left(x(w),x',s\right) dw~~~,
\label{2.2}
\end{equation}
where
$x(w)=\left(r\cos(\varphi +w),r\sin(\varphi+w)\right),
x'=\left(r'\cos \varphi',r'\sin \varphi '\right)$ and
\begin{equation}
K(x,x',s)={1 \over 4\pi s}
\exp \left(-{(x-x')^2 \over 4s}\right)~~~.
\label{2.3}
\end{equation}
The
contour $C$ in (\ref{2.2}) has two branches, one in the upper half complex
plane of the parameter $w$ going from $(-\pi -\triangle\varphi+i\infty)$ to
$(\pi-\triangle\varphi +i\infty)$ and the other in the lower half-plane
from $(\pi-\triangle\varphi-i\infty)$ to $(-\pi-\triangle\varphi-i\infty)$,
see Appendix A and Fig.1.b. in \cite{a12}.
It follows, in
particular, form the representation (\ref{2.2}) that $K_{\alpha}$ can also be
written as an infinite periodicity sum
\begin{equation}
K_{\alpha}(x,x',s)=\sum_{m=-\infty}^{\infty}K_{\infty}(x(m\alpha),x',s)
\label{2.4}
\end{equation}
of the heat kernel $K_{\infty}=\lim_{\alpha\rightarrow\infty} K_{\alpha}$ on
an infinitely-sheeted Riemann surface \cite{a5}.

It is useful to represent (\ref{2.2}) for $|\triangle\varphi|=
|\varphi-\varphi'|<\pi$ in a bit different form
\begin{equation}
K_{\alpha}\left(x,x',s\right)=K(x,x',s)+{i \over 2\alpha} \int_{\Gamma} \cot
\left(\pi \alpha^{-1} w\right) K\left(x(w),x',s\right) dw~~~,
\label{2.5}
\end{equation}
by explicitly writing the contribution of the plane heat kernel. In the
remaining
integral contour $\Gamma$ consists now of two curves, going from $(-\pi
-\triangle
\varphi +i\infty)$ to $(-\pi-\triangle\varphi-i\infty)$ and from $(\pi
-\triangle
\varphi -i\infty)$ to $(\pi-\triangle\varphi+i\infty)$ and intersecting the
real
axis between the poles of the integrand $-\alpha,~0$ and $0,~\alpha$
respectively. The equation (\ref{2.5}) can easily be obtained from
(\ref{2.2}) by transformating of the contour $C$.

Let us return now to the asymptotic expansion (\ref{3.1}) and calculate
the heat coefficients $a_{\alpha,n}(f)$.
It turns out that
$a_{\alpha,0}(f)$ is determined by the heat kernel on a plane due to the
first term in (\ref{2.5}) and simply is
\begin{equation}
a_{\alpha,0}(f)=\int \sqrt{g}~d^2xf(x)~~~,
\label{3.2}
\end{equation}
whereas the other coefficients $a_{\alpha,n}(f),~n\geq1,$ result from
the integral term in (\ref{2.5}).
To evaluate them, when the singularity at $r=0$ is taken into account,
we restrict the integration in (\ref{2.5}) by a
final part $\Gamma_R$ of the contour $\Gamma$ of a size $R$,
passing then from $\Gamma_R$ to $\Gamma$. One can thus write for the
difference
$$
\int\sqrt{g}~d^2x\left(K_{\alpha}(x,x,s)-K(x,x,s)\right)f(x)=
$$
\begin{equation}
=\lim_{\Gamma_R\rightarrow\Gamma} {i \over 8\pi \alpha s} \int_{0}^{\infty}rdr
f_0(r)\int_{\Gamma_R}\cot\left(\pi w \alpha^{-1}\right)
\exp\left(-{r^2\sin^2w/2
\over s}\right)~dw~~~,
\label{3.3}
\end{equation}
$f_0(r)\equiv\int_{0}^{\alpha}d\varphi~f(r,\varphi)$, and change the order
of integration. So far as $\Gamma$ can be chosen so that
$Re(\sin^2w/2)>0$, then as $s\rightarrow0$ the following
expansion
\begin{equation}
\int_{0}^{\infty}rdrf_0(r)\exp\left(-~{r^2\sin^2w/2 \over s}\right)= \frac 12
\sum_{n=o}^{\infty}{\Gamma\left((n+1)/2\right) \over n! (\sin^2w/2)^{(n+1)/2}}
{}~{d^n(rf_0(r)) \over dr^n}\vert_{r=0}~s^{{n+1 \over 2}}+ES
\label{3.4}
\end{equation}
holds ($\Gamma(x)$ denotes the gamma-function).
The sign in the square root of $\sin^2w/2$ in (\ref{3.4}) has
to be chosen from the conditions that are determined by the properties of the
integral over $r$
\begin{equation}
\left(\sin^2w/2\right)^{1/2}=\sin w/2,~~~~Re\sin w/2~>~0,
\label{3.5}
\end{equation}
$$
\left(\sin^2w/2\right)^{1/2}=-\sin w/2,~~~~Re\sin w/2~<~0,
$$
where the upper one is valid for the right part of $\Gamma$; whereas the
lower, for the left. By using (\ref{3.4}) it is not difficult to show now
that the action of the other functionals $a_{\alpha,n}$ on the considered
space of functions gives
\begin{equation}
a_{\alpha,n}(f)={\Gamma(n/2) \over (n-1)!}C_n(\alpha)\int_{0}^{\alpha}d\varphi
{}~{d^{n-1}\left(rf(r,\varphi)\right) \over dr^{n-1}}\vert_{r=0}~~,~~n\geq1~~~,
\label{3.6}
\end{equation}
where $C_n(\alpha)$ are the following integrals
\begin{equation}
C_n(\alpha)={i \over 4\alpha}\int_{\Gamma}\cot(\pi w\alpha^{-1})
(\sin^2w/2)^{-n/2} dw~~~.
\label{3.7}
\end{equation}
For $\alpha=2\pi$ the integrand in (\ref{3.7}) is a $2\pi$-periodic function
of $w$ and $\Gamma$ can be deformed so as the contributions of
both its parts to cancel each other. In this case all $C_n(\alpha)$
and consequently $a_{\alpha,n}(f)$ for $n\geq1$ turn out to be zero leaving
the only contribution in (\ref{3.7}) provided by the kernel $K(x,x',s)$.

It is important that according to (\ref{3.6}) the heat coefficients $a_{\alpha,
n}(f),~n\geq1$ act like a delta function and don't depend on the behavior of
$f$
at regular points on a cone. They would never appear if the integration over
the cone in (\ref{3.1}) were stopped short before the point $r=0$, by no
matter how close.

Integrals of the type (\ref{3.7}) have been discussed in
\cite{a13}. For even values of indices $n=2k$ they can be converted
to the following form
\begin{equation}
C_{2k}(\alpha)={i \over 4\alpha}\oint \cot(\pi w\alpha^{-1})
(\sin^2w/2)^{-2k} dw
\label{3.8}
\end{equation}
and represented in terms of polynomials of the order $2k$ in powers of
$\alpha^{-1}$.
We list here the values of the first two ones for $k=1,2$
\begin{equation}
C_2(\alpha)=\frac 16 \left(\left(2\pi \alpha^{-1}\right)^2-1\right)~~~,
\label{3.9}
\end{equation}
\begin{equation}
C_4(\alpha)={1 \over 90} \left(\left(2\pi \alpha^{-1}\right)^2-1\right)
\left(\left(2\pi \alpha^{-1}\right)^2+11\right)~~~
\label{3.10}
\end{equation}
to be required for the further analysis.
However, as for the odd indices, the quantities $C_{2k+1}(\alpha)$ can be
given only in an integral form, see \cite{a13}.

For a particular but important case when the function in (\ref{3.1}) is
assumed to be equal to unity in a domain of $V$ of the cone including its apex
and zero at other points the series is truncated and one gets the
expression exact up to the $ES$ terms
\begin{equation}
Tr\left(e^{-s\triangle_{\alpha}}f_V\right)={1 \over 4\pi s}
\left(V+\alpha C_2(\alpha)~s\right)+ES~~~.
\label{3.11}
\end{equation}
This result can be immediately checked for certain values $\alpha=2\pi n^
{-1}~~(n=2,3,...)$ when $K_{\alpha}(x,x',s)$ is explicitly presented as
a finite periodicity sum of $K(x,x',s)$.

It is worth also to point out that expression (\ref{3.1}) can trivially be
generalized to the heat kernel on the space product of a cone
and a smooth manifold. For instance, if the latter is the $d-2$-dimensional
Euclidean space $R^{d-2}$ with the Laplace operator $\triangle_{d-2}$,
one can write, by using (\ref{3.11}),
\begin{equation}
Tr\left(e^{-s(\triangle_{d-2}+\triangle_{\alpha})}\right)=
{1 \over (4\pi s)^{d/2}}
\left(\Omega_d+\Sigma_{d-2}\alpha C_2(\alpha)~s\right)+ES~~~,
\label{3.12}
\end{equation}
where $\Omega_d=\Sigma_{d-2}V$ is the volume of the total space and
effect of the conical singularities consists in appearing of the "surface" term
proportional to the volume $\Sigma_{d-2}$ of the hypersurface $r=0$.

So far as
the space is non-compact, $\Omega_d$ and $\Sigma_d$ are to be infinite and thus
(\ref{3.12}) has to be treated in a regularized sense like (\ref{3.11}). In
this case the $ES$ terms are significant. If $L$ is a typical size of the space
(the length at which the integrals are cut off), then $ES$ terms in
(\ref{3.12})
can be shown to be of the order $s^{-(d-2)/2}\exp(-L^2/s)$. From now on we
drop $ES$ as negligible in the limit $L\rightarrow\infty$ we are
interested in.

\section{Quantum field near cosmic string}
\setcounter{equation}0

Let us consider a quantum scalar field near a cosmic string being
in the flat space-time. For simplicity we confine the following
analysis to the case of an infinitely thin straight string that is at rest
along
the $z$ axis. The metric around it can be written in the form
\begin{equation}
ds^2=dt^2-dz^2-dr^2-r^2d\varphi ^2~~,~~~0\leq\varphi \leq\alpha
\label{5.1}
\end{equation}
and it is a solution of the Einstein equations \cite{a3},\cite{a4}
\begin{equation}
R_{\mu\nu}-\frac 12 g_{\mu\nu}R=-8\pi G T_{\mu\nu}
\label{5.2}
\end{equation}
where the energy-momentum tensor of the string, $T_{\mu\nu}$, has only two
non-zero components
\begin{equation}
T_{tt}=-T_{zz}=\mu\delta  _2(r)~~~,~~~\int_{0}^{\alpha}d\varphi
\int_{0}^{\infty} rdr~\delta_2(r)=1~~~.
\label{5.3}
\end{equation}
($\delta _2(r)$ can be represented with the help of the one-sided delta
function; $\delta _2(r)=$ \\ $=(\alpha r)^{-1}\delta (r+0)$.)
 From (\ref{5.2}) the string tension $\mu$ turns out to be immediately related
to the polar angle deficit \cite{a3}-\cite{a4}
\begin{equation}
\mu = {1 \over 8\pi G}(2\pi - \alpha)~~~.
\label{5.4}
\end{equation}
In this Section we concern the global effects of the vacuum polarization
on the string space-time (\ref{5.1}) that are displayed in the integral
quantities like the effective action $W$ or the ground energy $E_0(\alpha)$
of a quantum field around the string.

As for $E_0(\alpha)$, two different ways can be used to calculate this
quantity. The first one is to obtain $E_0(\alpha)$ as the integral of the
renormalized energy density $<{\hat T}_{00}(x)>^{\alpha}_{sub}$. However
this method cannot be applied immediately so far as the renormalized
energy momentum tensor has a non-integrable infinity at the string axis
\cite{a13}-\cite{a19} and an additional regularization will be shown
in the next Section to be needed. Here we consider another way based on
the thermodynamical relation between the internal energy $E_{\beta ^{-1}}$
of the system at a temperature $\beta ^{-1}$ and the partition function
$Z_{\beta}$
\begin{equation}
E_{\beta ^{-1}}=<{\hat H}>_{\beta}=-{\partial \over \partial \beta}
\log Z_{\beta}
\label{5.5}
\end{equation}
where ${\hat H}$ is the Hamiltonian.
In such approach the ground energy $E_0$ is the energy at zero temperature
$E_0=\lim_{\beta \rightarrow \infty} E_{\beta ^{-1}}$ and to get it in the
one-loop approximation the equation (\ref{3.12}) for the trace can be used.

The partition function is known to be represented in the form of a
functional integral by passing to an imaginary time. In particular,
for a self-interacting scalar field $\phi$ around the string
with a potential $V(\phi)$ one has
\begin{equation}
Z_{\beta}=Tr(e^{-\beta\hat{H}})=\int~D\phi e^{-S_e[\phi]}~~~,
\label{5.6}
\end{equation}
where $D\phi$ is an integration measure
and the action in the exponential
\begin{equation}
S_e[\phi]=\int\sqrt{g}d^4x \left(\frac 12 \partial _{\mu}\phi
\partial ^{\mu}\phi +V(\phi)\right)
\label{5.7}
\end{equation}
is given on the Euclidean
section of the space-time around the string
\begin{equation}
ds^2=d\tau ^2+dz^2+dr^2+r^2d\varphi ^2~~,~~~0\leq\varphi \leq\alpha~,~~
0\leq\tau\leq\beta~~~
\label{5.8}
\end{equation}
with periodicity in $\tau$.

Another quantity we are interested in is the effective action $W$ that
can be also defined for a finite temperature with the help of the partition
function
\begin{equation}
W=-\log Z_{\beta}~~~,
\end{equation}
by taking next the limit $\beta \rightarrow \infty$. Its variations coincide
with the thermal average of the functional $\delta S_e$ that is interpreted
as a quantum operator
\begin{equation}
\delta W=
Z^{-1}_{\beta}\int D\phi~ \delta S_e[\phi] e^{-S_e[\phi]}=
<\delta {\hat S}_e>_{\beta}~~~.
\end{equation}
Strictly speaking, $W$ is an Euclidean form of the effective action
but transition to the convenient definition \cite{a23},\cite{a2} doesn't
make any difficulty for the static space like (\ref{5.1}).

To obtain $W$ and $E_0(\alpha)$ in the one loop approximation let us
consider the system in a large but finite volume $\Omega _4$ of the space
(\ref{5.8}) that includes the string and expand the action $S_e$ near its
minimum $\varphi$ up to the second order terms
$$
S_e[\varphi + \phi ']=S_e[\varphi]+\frac 12 \int \sqrt{g} d^4x
\phi '[\triangle + M^2]\phi '
$$
$$
M^2\equiv V''(\varphi)
$$
($\triangle$ denotes the Laplace operator on (\ref{5.8})). Assuming $\varphi$
to be a constant configuration and by integrating in (\ref{5.6}) over $\phi '$
one gets
\begin{equation}
W[\varphi ]
=\Omega _4~V(\varphi)+ {\hbar \over 2}\log\det(\triangle +
M^2) +O(\hbar ^2) ~~~,
\label{5.13}
\end{equation}
\begin{equation}
E_0={\partial \over \partial \beta} W~~~~~(\beta \rightarrow \infty)
\label{5.14}
\end{equation}
where $\Omega _4 =\beta \int dv$ and
the Planck constant $\hbar$ is introduced explicitly to
emphasize the quantum corrections.

The second term in (\ref{5.13}) is ultraviolet divergent and to get a finite
expression we have to renormalize the effective action $W[\varphi]$.
To this end the dimensional regularization \cite{a15}, for instance, can be
used. It
suggests in our case that the space (\ref{5.8}) has to be changed to
the space product
$R^{d-2} \otimes Cone$, passing to arbitrary values of the parameter $d$.
The quantity $\log \det(\triangle +M^2)$ can be evaluated then with the help of
the following representation \cite{a16}
\begin{equation}
\log \det(\triangle + M^2)=
Tr\log(\triangle + M^2)=-\int_{0}^{\infty}{ds \over s}
Tr\left(e^{-s\triangle}\right)e^{-M^2s}~~~,
\label{5.15}
\end{equation}
where at low temperature $(\beta\rightarrow\infty)$
the equation (\ref{3.12}) for the trace of the heat kernel on
$R^{d-2}\otimes Cone$ is valid. Besides, in (\ref{3.12}) the
"regularized" volumes $\Omega _d$, and $\Sigma _{d-2}$ are
expressed through the physical volume $\Omega _4=\beta\int dv$ and the area
$\Sigma _2=\beta\int dz$ of the surface $r=0$
\begin{equation}
\Omega _{d}=\nu ^{\epsilon}\Omega _4~~,~~~\Sigma _{d-2}=
\nu ^{\epsilon} \Sigma _2~~~,
{}~~~\epsilon \equiv 4-d~~~.
\label{5.16}
\end{equation}
Here an additional parameter $\nu$ with the mass dimension
is introduced to adjust the dimensions of the left and right sides
of these equalities. After the integration in (\ref{5.15})
the regularized $\log \det(\triangle +M^2)$ at $\beta \rightarrow \infty$
reads
$$
\log \det(\triangle +M^2)=- \Omega _4\left({4\pi \nu ^2 \over M^2}\right)^
{\epsilon /2} {M^4 \over 16 \pi ^2} \Gamma (\epsilon /2 -2)
$$
\begin{equation}
-\Sigma_2 \left({4\pi \nu ^2 \over M^2}\right)^{\epsilon /2}
{M^2 \over 16 \pi ^2} \alpha C_2(\alpha) \Gamma(\epsilon /2 -1)~~~.
\label{5.17}
\end{equation}

 From (\ref{5.13}) and  (\ref{5.17})
one can see that in our case as distinct from the theory
in the Minkowsky space an additional surface term proportional to $\Sigma _2$
appears in the effective action $W$.
After passing from the metric (\ref{5.8}) to (\ref{5.1}) the new term in $W$
is represented by an integral over the string world sheet. Consequently
it is worth to unify it with the string action $\mu \Sigma _2$ that can be
added to $W$. This gives rise to a surface effective action in the total
functional.

To investigate now the renormalization let us consider a simple model of a real
scalar field with the potential
\begin{equation}
V(\varphi)={m^2 \over 2}\varphi ^2 + {\lambda \over 24} \varphi ^4~~~,~~~~m^2~,
\lambda >0~~~.
\label{5.18}
\end{equation}
In this case the total regularized one-loop
effective action expressed through the bare parameters $\lambda_B,m_B,\mu_B$
looks as follows
$$
W_{tot}[\varphi]=W[\varphi]+W_{B,surf}[\varphi]=
$$
\begin{equation}
=\Omega _d\left({m_B^2 \over 2}\varphi _B ^2 +
{\lambda _B \over 24} \varphi _B^4\right)+
{\hbar \over 2}\log\det (\triangle +M^2)+
O(\hbar ^2)+
\Sigma_{d-2}\left(\mu _B+\sigma _B \varphi _B^2 \right)
\label{5.19}
\end{equation}
where apart from the bare string action an additional term $\sigma _B\varphi
_B^2$ is included in the bare surface functional $W_{B,surf}=\Sigma _{d-2}
(\mu_B +\sigma _B\varphi _B^2)$ to eliminate the corresponding divergence.
To remove
$$
{}~~~~~~~~~the~pole~part~of~~\log\det(\triangle +M^2)=
$$
$$
={1 \over \epsilon}\left[-\Omega _4{1 \over 16\pi ^2}(m^2+\lambda \varphi ^2/2)
^2+\Sigma_2 {\alpha C_2(\alpha) \over 16 \pi ^2}(m^2+\lambda \varphi ^2/2)
\right]~~~,
$$
from the functional $W[\varphi]$ the bare constants
$\lambda _B, m_B^2, \mu_B, \sigma _B$ have to be expressed
through the renormalized ones $\lambda, m^2, \mu, \sigma$
\begin{equation}
\lambda _B=\nu ^{\epsilon}\left(\lambda+{\hbar \over \epsilon}
{3\lambda ^2 \over 16 \pi ^2} + O(\hbar ^2)\right)~~~,
\label{5.20}
\end{equation}
\begin{equation}
m^2_B=m^2\left(1+{\hbar \over \epsilon}
{\lambda \over 16 \pi ^2} + O(\hbar ^2)\right)~~~,
\label{5.21}
\end{equation}
\begin{equation}
\mu _B=\nu ^{-\epsilon}\left(\mu-{\hbar \over \epsilon}
{m^2\alpha C_2(\alpha) \over 16 \pi ^2} + O(\hbar ^2)\right)~~~,
\label{5.22}
\end{equation}
\begin{equation}
\sigma _B=\sigma-{\hbar \over \epsilon}
{\lambda ^2\alpha C_2(\alpha) \over 32 \pi ^2} + O(\hbar ^2)~~~.
\label{5.23}
\end{equation}
The above definitions correspond to a renormalization
recipe in which the finite parts of the counterterms are assumed to be equal to
zero \cite{a15}. As for the bare field $\varphi _B$, it is related to
the renormalized one $\varphi$ by the equality $\varphi _B= \nu^{-\epsilon /2}
\varphi$ because any counterterms in $\varphi _B$ can always be removed,
shifting the variable of integration in (\ref{5.6}). Differentiating the
equations (\ref{5.20})-(\ref{5.23}) with respect to the parameter $\nu$
it is easy
to get apart from the standard renormgroup equations for $\lambda$ and $m^2$
the new ones for the string tension $\mu$ and $\sigma$.

The total one-loop effective action $W_{tot}$ written in terms of the
renormalized quantities
defined
by (\ref{5.20})-(\ref{5.23}) can be represented as a sum of the volume and
surface parts
\begin{equation}
W_{tot}[\varphi]=W_{vol}[\varphi]+W_{surf}[\varphi]
=\Omega _4~V_{eff}(\varphi) + \Sigma _2~ \mu (\varphi)
\label{5.25}
\end{equation}
where for a constant argument the functional $W_{vol}[\varphi]$ is expressed
through
the renormalized effective potential of the system $V_{eff}[\varphi]$
that looks the same as in the Minkowsky space.
This fact can be used to fix the value of the
renormalization parameter $\nu$.
For instance, in the considered model one can identify $m$ with the physical
mass, that is equivalent to the following normalization condition \cite{c1}
\begin{equation}
V''_{eff}(0)=m^2~~~
\label{5.24}
\end{equation}
at the minimum $\varphi =0$ of $V_{eff}$. It gives the relation
$4\pi \nu ^2=m^2\exp(1/2+\gamma)$, where $\gamma$ is the Euler constant, and
\begin{equation}
V_{eff}(\varphi)={m^2 \over 2}\varphi ^2 +{\lambda \over 24} \varphi ^4 +
{\hbar \over 64 \pi ^2}\left(m^2+\lambda \varphi ^2/2\right)^2
\left(\log\left({m^2+\lambda \varphi ^2/2 \over m^2}\right)-\frac 12\right)~~~.
\label{5.26}
\end{equation}
Besides, it results to the renormalized surface effective action that
can be represented as follows
\begin{equation}
W_{surf}[\varphi]=\Sigma _2\mu(\varphi)=\Sigma _2\left[\mu +\sigma \varphi ^2-
{\hbar \over 32 \pi ^2}\alpha C_2(\alpha) \left(m^2+\lambda \varphi ^2/2\right)
\left(\log\left({m^2+\lambda \varphi ^2/2 \over m^2}\right)-\frac 12\right)
\right]~~~.
\label{5.27}
\end{equation}

Finally, the total renormalized energy of the string and quantum field can be
obtained in accordance with (\ref{5.14}) replacing there the functional
$W[\varphi]$ to the total renormalized effective action $W_{tot}[\varphi]$.
This is the same as if we changed the definition (\ref{5.6}) of the partition
function $Z_{\beta}$ and added to the functional $S_e[\phi]$ the surface
action $S_{surf}[\varphi]=\int d\tau dz(\mu+\sigma \phi ^2)$. Thus, after
subtracting
the vacuum energy $E_{0,Mink}=\int dv V_{eff}(0)$ in the Minkowsky space,
we come to result
\begin{equation}
E_{tot}={\partial \over \partial \beta}W_{tot}[0]-E_{0,Mink}
=\mu (0)\int dz~~~.
\label{5.28}
\end{equation}
This quantity is taken at the minimum $\varphi =0$ of the potential $V_{eff}
(\varphi)$ that corresponds to a field configuration with zero average field
strength $<{\hat \phi}>=0$ \cite{c1}. It follows from (\ref{5.28}) that
renormalized
energy per unite length turns out to be finite and equal to
\begin{equation}
\mu (0)=\mu+{\hbar \over 64\pi ^2}m^2\alpha C_2(\alpha)
+O(\hbar ^2)~~~.
\label{5.29}
\end{equation}
So far as $\mu(0)$ occurs from the surface functional (\ref{5.27}) the non-zero
value of $E_{tot}$ is completely provided by the energy density at the string
axis.

The constant $\mu(0)$ should be considered as an effective tension of the
string that includes quantum corrections to the classical tension $\mu$
related with the parameter $\alpha$ by (\ref{5.4}). It is interesting that
the renormalized surface action $W_{surf}[\varphi]$ depends on $\varphi$
even if $\sigma =0$ and therefore in general case the effective tension
$\mu(\varphi)$ varies if the average value of the field $<{\hat \phi}>=\varphi$
changes that happens in the case of a phase transition.

\section{Total energy end energy density}
\setcounter{equation}0

Until now we dealt with the integral quantities like
the effective action and total energy using for their
calculation the trace of the heat kernel. The surface terms appearing in these
quantities have a global origin: they would have not arisen, if we had
excluded,
from the integrals over the space-time, the region around the string world
sheet. The local renormalized energy momentum tensor near the cosmic string
was calculated by a number of authors \cite{a13}-\cite{a19}.
Let us find out a connection between their and our results and demonstrate
that the local non-integrable divergence in the average energy density
arising as the string is approached can be removed by a suitable
renormalization of the bare string tension so that the total energy
turns out to be finite. What we are going to do is to explore the same
approach as used in \cite{a20} in quantum theory with boundaries.

We consider a real massless scalar field for which the energy density can be
obtained in the closed form \cite{a13}
\begin{equation}
<{\hat T_{00}}(x)>^{\alpha}_{sub}={1 \over 16\pi ^2 r^4}\left(2(1-4 \xi )
C_2(\alpha)-C_4(\alpha)\right)
\label{6.1}
\end{equation}
expressed in terms of the polynomials (\ref{3.9}),(\ref{3.10}). The value
$\xi =1/6$ corresponds to a conformally invariant field. The local energy is
evaluated in a standard way from the Green function $G^{\alpha}(x,x')=
i^{-1}<T({\hat \phi}(x),{\hat \phi}(x'))>$ as a coincidence limit
\begin{equation}
<{\hat T_{00}}(x)>^{\alpha}_{sub}=\lim_{x'\rightarrow x}{\cal T}_{00}
G^{\alpha}_{sub}(x,x')
\label{6.2}
\end{equation}
where ${\cal T}_{00}$ denotes a second order differential operator \cite{a21}
depending on the type of field and the divergences are removed by subtracting
from the Minkowsky Green function $G^{\alpha =2\pi}$ from $G^{\alpha}$
\begin{equation}
G^{\alpha}_{sub}(x,x')=G^{\alpha}(x,x')-G^{\alpha =2\pi}(x,x')~~~.
\label{6.3}
\end{equation}

It is obvious that the local divergence of
the energy density (\ref{6.1}) at $r=0$ can be
regularized if we restrict the domain of integration in the total energy
by the values of coordinates $r\geq r_0$ where $r_0$ is a positive small
parameter that can be treated as the string radius.
Moreover the regularization suggested
also makes finite the surface term in the effective
action. To see this it is worth to use the equation (\ref{2.5}), which gives,
instead of (\ref{3.12}),
$$
Tr\left(e^{-s(\triangle_{\alpha}+\triangle _2)}\right)_{r_0}\equiv
\int_{r_0}^{\infty}rdr \int_{0}^{\alpha}d\varphi~K_{\alpha}(x,x,s)Tr\left(e^{-s
\triangle}\right)=
$$
\begin{equation}
={1 \over (4\pi s)^2}\left(\Omega _4+\Sigma _2 {is \over 4}\int_{\Gamma}
{\cot (\pi \alpha ^{-1}w) \over \sin ^2w/2}\exp\left(-{r_0^2\sin^2w/2 \over s}
\right)dw
\right)~~~.
\label{6.4}
\end{equation}
Then for a free scalar field the total one-loop effective action (in the case
when $\sigma _B=0$)
can be defined like (\ref{5.19}) and takes
the form
$$
W_{tot}={\hbar \over 2}\log\det(\triangle +m^2)_{r_0}+\mu _B\Sigma _2
=-{\hbar \over 2}\int_{0}^{\infty}{ds \over s}Tr\left(e^{-s(\triangle _{\alpha}
+\triangle _2)}\right)_{r_0}e^{-m^2 s}+\mu_B\Sigma _2 =
$$
\begin{equation}
\equiv W_{vol}+W_{r_0,surf}
\label{6.5}
\end{equation}
It is separated into the volume part $W_{vol}$ proportional to $\Omega _4$
and the surface part $W_{r_0,surf}$ given on the world sheet $\Sigma_2$. As
distinct from $W_{vol}$ developing the standard divergences, the surface
action $W_{r_0,surf}$ now turns out to be finite while $r_0\neq 0$ and in the
massless case
$m^2=0$ its expression results from (\ref{6.4})
$$
W_{r_0,surf}=\Sigma _2 \left(\mu_B-{i\hbar \over 8}\int_{0}^{\infty}
{ds \over (4\pi s)^2} \int_{\Gamma}
{\cot (\pi \alpha ^{-1}w) \over \sin ^2w/2}\exp\left(-{r_0^2\sin^2w/2 \over s}
\right)dw\right)=
$$
\begin{equation}
=\left(\mu _B -\hbar {\alpha C_4(\alpha) \over 32\pi ^2r_0^2}\right)\Sigma
_2~~~.
\label{6.6}
\end{equation}
It follows from (\ref{6.6}) that the divergence in $W_{r_0,surf}$ at
$r_0\rightarrow 0$ can be
removed by replacing as before the bare string tension $\mu_B$ by the
renormalized $\mu$
\begin{equation}
\mu _B=\mu+{\hbar \over 32\pi ^2r_0 ^2}\alpha C_4(\alpha)~~~.
\label{6.7}
\end{equation}
Taking this into account one can write the local renormalized energy as the
sum
\begin{equation}
<{\hat T}_{00}>^{\alpha}_{r_0,ren}=T_{00,B}+<{\hat T}_{00}>^{\alpha}_{r_0,sub}
\label{6.8}
\end{equation}
of the string energy $T_{00,B}=\mu _B\delta _2(r)$ concentrated at the string
axis and the renormalized energy density of the quantum field in the domain
$r\geq r_0$. Two densities, $<{\hat T}_{00}>^{\alpha}_{sub}$ given by
(\ref{6.1}) and
$<{\hat T}_{00}>^{\alpha}_{r_0,sub}$ coincide everywhere except the region near
the string. To demonstrate this let us calculate the classical energy-momentum
tensor of the field in this domain defined by the functional differentiation
of the action that we take in the same form as in \cite{a20}
\begin{equation}
S=-\frac 12 \int_{r\geq r_0} d^4x \sqrt{-g}~ \phi(x)[\Box +\xi R]\phi(x)
\label{6.9}
\end{equation}
where $\Box =\sqrt{-g}~^{-1}\partial_{\mu}\sqrt{-g}g^{\mu\nu}\partial_{\nu}$
is the D'Alambertian and $R$ is the scalar curvature. The variation of this
functional $\delta S$ under changing the metric $\delta g^{\mu\nu}$
consists of two parts
\begin{equation}
\delta S=\frac 12 \int_{r\geq r_0} d^4x \sqrt{-g}~T_{\mu\nu}(x)\delta
g^{\mu\nu}(x)+\delta _{surf}S
\label{6.10}
\end{equation}
where $T_{\mu\nu}$ stands for the normal expression of the stress tensor of a
scalar field \cite{a16} and an additional surface term arises due to the
restriction of the domain of integration
$$
\delta _{surf}S=-\frac 12 \int_{r=r_0} d\sigma ^{\tau} \left[ \phi
^2(g_{\mu\nu}
\delta g^{\mu\nu}~_{;\tau} + g_{\tau\mu}\delta g^{\mu\sigma}~_{;\sigma})\right.
$$
\begin{equation}
+\left.\left((1/4-\xi)(\phi ^2),_{\tau}g_{\mu\nu}+(\xi-1/2)(\phi^2),_{\nu}
g_{\mu\tau}\right)\delta g^{\mu\nu}\right]
\label{6.11}
\end{equation}
($d\sigma ^{\tau}$ is the area element). So far as there is no real boundary
of the space on the surface $r=r_0$, the variations of the metric
$\delta g^{\mu\nu}|_{r=r_0}$ don't vanish on it. They are independent
of its normal derivatives on the surface and thus the last ones
can be ignored. As a result, $\delta _{surf}S$ produces the additional term
in the energy density
\begin{equation}
T_{00,surf}={2 \over \sqrt{-g}}{\delta _{surf}S \over \delta g^{00}}
=(1/4-\xi)\delta(r-r_0){d \over dr}(\phi)^2
\label{6.12}
\end{equation}
giving rise to the distinction between the average density in the domain,
$<{\hat T}_{00}>^{\alpha}_{r_0,sub}$, and the local energy (\ref{6.1})
\begin{equation}
<{\hat T}_{00}>^{\alpha}_{r_0,sub}=<{\hat T}_{00}>^{\alpha}_{sub}+
i(1/4 -\xi)\delta (r-r_0)\lim_{x\rightarrow x'}\left({d \over dr}
+{d \over dr'}\right)G^{\alpha}_{sub}(x,x')
\label{6.13}
\end{equation}
($\delta (r-r_0)$ is the one-sided delta-function). For its calculation
the proper-time representation for the Green function \cite{a1},\cite{a2}
written in the form
$$
G=-(\Box+m^2)^{-1}=-\int_{0}^{\infty}ds ~e^{-(\Box+m^2)s}
$$
can be used. It gives, together with (\ref{2.5}), the subtracted Green function
at $t=t',z=z'$ and $\varphi=\varphi '$ by the integral
\begin{equation}
G^{\alpha}_{sub}(r,r')=-{i \over 16\pi ^2}\int_{0}^{\infty}{ds \over s^2}
{i \over 2\alpha}\int_{\Gamma}\cot(\pi \alpha ^{-1}w)\exp\left(-
{r^2+r'^2-2rr'\cos w \over 4s}\right)dw
\label{6.14}
\end{equation}
which can be substituted into (\ref{6.13}) to obtain
\begin{equation}
<{\hat T}_{00}>^{\alpha}_{r_0,sub}=<{\hat T}_{00}>^{\alpha}_{sub}-
(1/4 -\xi){C_2(\alpha) \over 4\pi r_0^3}\delta(r-r_0)~~~.
\label{6.15}
\end{equation}
Integrating now the renormalized quantity (\ref{6.8}) over the space
\begin{equation}
E_{tot}=\int<{\hat T}_{00}>^{\alpha}_{r_0,ren}~dv =
\left[\mu _B+\int_{r_0}^{\infty}rdr\int_{0}^{\alpha}d\varphi<{\hat T}_{00}>
_{r_0,sub}^{\alpha}\right]\int dz
\label{6.16}
\end{equation}
and using (\ref{6.7}),(\ref{6.15}) we find that the counterterm in the bare
tension $\mu _B$ cancels exactly the term proportional to $r_0 ^{-2}$ in the
integrated energy of the field rendering finite
the renormalized total energy at
$r_0\rightarrow 0$
\begin{equation}
E_{tot}=\mu\int dz~~~.
\label{6.17}
\end{equation}
This shows explicitly that finiteness of the total energy derived in the
previous Section is a consequence of renormalization of the bare string
tension. There is
also quantitative agreement between (\ref{6.17}) and equation (\ref{5.28})
where for zero mass $\mu(0)= \mu$ and in both cases the parameter $\mu$ has to
be
identified with the classical string tension.

\section{Conclusions and remarks}

In this work a close analogy between quantum theory on the space with
conical singularities and quantum theory with boundaries was outlined.
In both cases the one loop quantum corrections result to divergent surface
functionals in the effective actions. The renormalization of these functionals
can be used to remove non-integrable divergence in the energy density and
to obtain the finite total energy of the system.
However, this analysis
concerns the idealized objects, strings and boundaries of zero
thickness.
In effect one might expect that for the real string of a finite
size the divergent terms on its world sheet give large but still finite
contributions to the renormalized energy.

In the theory with boundaries the surface actions are known to essentially
depend on which of the boundary conditions, Dirichlet or Neumann, are imposed.
As for the string case, we used the finite boundary condition on the string
axis taken in Section 2 and others possibilities are worth to be investigated
as well. For example, the possibility of logarithmically divergent
conditions has been pointed out in \cite{a11} in connection with the
self-adjoint extensions of the Laplace operator on a cone. A hypothesis
has been made there that effects of the true interaction of the cosmic
string with the field can be taken into account by choosing one of the suitable
extensions provided we are interested in what happens in sufficiently large
length scales.

It is to be also mentioned that our consideration was virtually confined
to conical singularities in the flat space and incorporation of the
curvature effects represents an interesting problem.

{\bf Note added.} When this work were in press a preprint \cite{a25} was
appeared where scalar fields in the presence of conical singularities
were analyzed also and the same equation as the key equation (\ref{3.11})
for the trace of the cone heat kernel was obtained.

\bigskip

{}~~~~~~~~~~{\bf Acknowledgements}

\bigskip

This work was supported by the Heisenbebg-Landau program, Project N 30.
The author would also like to thank Professors V.G.Kadyshevsky and D.I.Kazakov
for encouragement and interest in his work and Dr. S.N.Solodukhin for
helpful discussions.

\newpage

{\appendix \noindent{\large \bf Appendix A. The zeta-function on a cone and
on a sphere with conical singularities}}\\
\def\theequation{A.\arabic{equation}}
\setcounter{equation}0

The generalized zeta-function on a cone can also be considered as a
functional on the chosen space of functions and
related with the heat kernel via the Mellin transform \cite{a14}
\begin{equation}
\zeta_{\alpha}(z,f)={1 \over \Gamma(z)}\int_{0}^{\infty}s^{z-1}e^{-m^2 s}
Tr\left(e^{-s\triangle_{\alpha}}f\right)~~~,
\label{4.1}
\end{equation}
with a mass $m$ providing convergence of the integral as
$s\rightarrow\infty$. As the cone is a non-compact space, the convenient
zeta-function, that is introduced through the
trace of the heat kernel \cite{a14}, can be defined as
$\zeta_{\alpha}(z,f_V)$
with the help of the function $f_V$ used in Section 2.
It follows immediately from (\ref{3.11}) that
\begin{equation}
\zeta_{\alpha}(z,f_V)={1 \over 4\pi}\left({V (m^2)^{1-z} \over z-1} +
a_{\alpha,2}(f_V) (m^2)^{-z}\right)~~~.
\label{4.3}
\end{equation}
In particular,
taking into consideration (\ref{3.9}) one gets from (\ref{4.3}) at
$m=0$ the finite expression that doesn't depend on the volume $V$
\begin{equation}
\zeta_{\alpha}(0)={\alpha \over 24\pi}\left( \left({2\pi \over \alpha
}\right)^2
-1\right)~~~.
\label{4.4}
\end{equation}

To take in our analysis the effects of curvature of the space, although
simple ones,
it is worth to compare $\zeta_{\alpha}(0)$ at $m=0$ with the
zeta-function of the
Laplace operator on the unit "sphere" with two conical singularities at
"south" and "north" poles, where the corresponding line element reads
\begin{equation}
ds^2=\cos^2\chi
d\varphi^2+d\chi^2,~~~~0\leq\varphi\leq\alpha,~~~|\chi|\leq\pi/2
\label{4.5}
\end{equation}
and takes the form (\ref{1.1}) as $|\chi|\rightarrow\pi/2$. Everywhere
at the other points the metric (\ref{4.5}) is regular and the space
has a finite constant curvature, the same as the curvature of the ordinary
unit sphere.

This example is interesting since the spectrum
of the Laplace-Beltrami operator on (\ref{4.5}) can be calculated exactly.
It is determined by two non-negative integers $n$ and $m$
\begin{equation}
\lambda_{n,m}=\left(n+(2\pi\alpha^{-1})m\right)\left(n+(2\pi\alpha^{-1})m +1
\right)
\label{4.6}
\end{equation}
with the double degeneracy for $m\neq 0$. Using the same transformations
that were carried out in \cite{a6} for the case of four-dimensional
analog of this "sphere", one can represent
the zeta-function for $z\rightarrow 0,-1,-2,..$
by the series
$$
\zeta^{sphere}_{\alpha}(z)=\sum_{n,m}\lambda_{n,m}^{-z}=
$$
\begin{equation}
={\alpha \over \pi}\sum_{k=0}^{\infty}\sum_{n=0}^{\infty} {\Gamma(z+k)
\Gamma(2z+2k+2n-1) \over 2^{2k}k!~\Gamma(z)\Gamma(2z+2k)}~
\zeta_R(2z+2k+2n-1,1/2)~{B_{2n} \over (2n)!}\left({2\pi \over \alpha}\right)
^{2n}
\label{4.7}
\end{equation}
where $\zeta_R$ is the Riemannian zeta-function and $B_{2n}$ are the Bernulli
numbers. It is not difficult to show in particular that for $z=0$ it is
given by the simple expression
\begin{equation}
\zeta_{\alpha}^{sphere}(0)={\alpha \over 6\pi} \left[1 +\frac 12
\left( \left({2\pi \over \alpha}\right)^2-1\right)\right]~~~.
\label{4.8}
\end{equation}
Apart from the contribution \cite{a14}
$$
{\alpha \over 6\pi}=-{1 \over 4\pi}\int \sqrt{g}d^2x~{R \over 6},
$$
determined at the points where the metric (\ref{4.5}) is regular
by the scalar curvature of the sphere $R=-2$, it contains also an
additional term, appearing because of the conical singularities at
$|\chi|=\pi/2$ and equal exactly to the doubled value of the zeta-function on
a cone $2\zeta_{\alpha}(0)$.

This result could be anticipated in advance, by taking into account that near
each of the points $\chi=\pm \pi/2$ the heat kernel expansion on the space
(\ref{4.5}) can be approximated by the expansion on a cone (\ref{3.1}).

\end{document}